\documentclass[namedreferences]{SSRv}

\usepackage{graphicx}
\usepackage{mathtime}
\usepackage{url}

\usepackage{rotating}

\newcommand{\QI}{\,Q_{\mathrm{I}}}

\begin{document}
\begin{article}
\begin{opening}

\title{Interstellar Dust in the Solar System}
\runningtitle{Interstellar Dust in the Solar System}

\author{Harald \surname{Kr\"uger}$^{1,2\ast}$}
\author{Markus \surname{Landgraf}$^{3}$}
\author{Nicolas \surname{Altobelli}$^4$}
\author{Eberhard \surname{Gr\"un}$^{2,5}$}
\institute{
$^1$Max-Planck-Institut f\"ur Sonnensystemforschung, 37191 Katlenburg-Lindau, Germany \\
$^2$Max-Planck-Institut f\"ur Kernphysik, 69029 Heidelberg, Germany \\
$^3$European Space Agency, ESOC, 64293 Darmstadt, Germany \\
$^4$NASA/JPL, Pasadena, California, USA \\
$^5$HIGP, University of Hawaii, Honolulu, HI 96822, USA \\
($^{\ast}$Author for correspondence, Email: krueger (at) mps.mpg.de)}

\runningauthor{H. Kr\"uger et al.}

\received{} \revised{} \accepted{}

\begin{abstract}
The Ulysses spacecraft has been orbiting the Sun on a highly inclined 
ellipse almost perpendicular to the ecliptic plane 
(inclination $79^{\circ}$, perihelion distance 1.3 AU, aphelion distance 5.4 AU) 
since it encountered Jupiter in 1992. The in-situ dust detector
on board continuously measured interstellar dust grains with masses up
to $\rm 10^{-13}$\,kg, penetrating deep into the solar system.
The flow direction is close to the mean apex of the Sun's motion through
the solar system and the grains act as tracers of the physical conditions 
in the local interstellar cloud (LIC). 
While Ulysses monitored the interstellar dust stream at high ecliptic 
latitudes between 3 and 5~AU, interstellar impactors were also measured 
with the in-situ dust detectors on board Cassini, Galileo and Helios, 
covering a heliocentric distance range between 0.3 and 3 AU in the 
ecliptic plane. The interstellar dust stream in the inner solar system
is altered by the solar radiation pressure force, gravitational focussing and
interaction of charged grains with the time varying interplanetary 
magnetic field.
We review the results from in-situ interstellar dust measurements 
in the solar system
and present Ulysses' latest interstellar dust data.
These data indicate a $30^{\circ}$ shift in the impact direction of 
interstellar grains w.r.t. the interstellar helium flow direction, the
reason of which is presently unknown.
\end{abstract}


\end{opening}

\section{Introduction}

One of the most important results of the Ulysses mission is the 
identification and characterization of a wide range of interstellar 
phenomena inside the solar system. A surprise was the identification
of interstellar dust grains sweeping through 
the solar system \cite{gruen1993}. Before this discovery it was believed that
interstellar grains  are prevented from reaching the planetary region
by electromagnetic  interaction with the solar wind magnetic field.
The interplanetary zodiacal  dust flux was thought to dominate the
near-ecliptic planetary region while  at high ecliptic latitudes only
a very low flux of dust released from  long-period comets should be
present. Therefore, the characterization of  the interplanetary dust
cloud was the prime goal of the Ulysses dust investigation \cite{gruen1992b}. 

\begin{figure}
\hspace{-0.4cm}
\centerline{\includegraphics[clip=,width=1.21\textwidth]{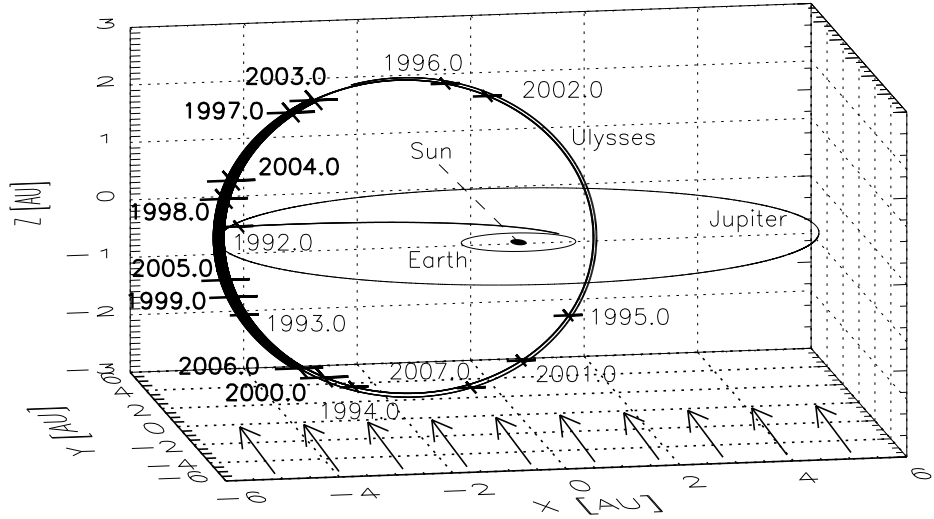}}
        \caption{\label{trajectory}
The trajectory of Ulysses in ecliptic coordinates. The Sun is in the centre. 
The orbits of Earth and Jupiter indicate the ecliptic plane. Ulysses' initial
trajectory was in the ecliptic plane. Since Jupiter flyby 
in early 1992 the orbit has been almost perpendicular to the ecliptic plane 
(79$^{\circ}$ inclination).
Crosses mark the spacecraft position at the beginning 
of each year. The 1997 to 1999 and 2003 to 2005 parts of the trajectory are 
shown as a thick line. Vernal equinox is to the right (positive x axis)
and the flow direction of the interstellar grains (coincident with the
interstellar helium flow) is indicated by arrows.
}
\end{figure}

Ulysses was launched in October 1990. A swing-by manoeuvre
at Jupiter in February 1992 rotated its orbital plane 
$ 79^{\circ}$ relative to the ecliptic plane (with a six-year orbital period 
and aphelion distance at 5.4~AU; Figure~\ref{trajectory}). Subsequent aphelion 
passages occurred in April 1998 and in June 2004. A second
Jupiter flyby occurred in February 2004. The best 
conditions for detection of 
interstellar impactors are in the outer solar system beyond 3~AU at high 
ecliptic latitudes and far away 
from Jupiter where impact rates of interplanetary grains or jovian stream 
particles are comparatively small. Nevertheless, 
Ulysses has continuously monitored the interstellar dust flux in the 
heliosphere since 1992. 

We briefly review the results from in-situ interstellar dust measurements 
with the Ulysses and other space-borne dust detectors in Section~\ref{sec_review}. 
In Section~\ref{sec_meas} we present the interstellar dust measurements from 
the Ulysses' $\mathrm{3^{rd}}$ passage through the outer heliosphere, 
and Section~\ref{sec_discussion} is a brief discussion of our results. 

\section{Interstellar dust measurements in the heliosphere}

\label{sec_review}

Interstellar dust particles originating from the Local Interstellar Cloud (LIC) 
move on hyperbolic 
trajectories through the solar system and approach Ulysses predominantly 
from the direction opposite to the expected impact direction
of interplanetary grains. On average, their impact velocities 
exceed the local solar system escape velocity, even if 
radiation pressure effects are neglected \cite{gruen1994a}. 
The grain motion through the solar system was found to be parallel to the 
flow of neutral interstellar hydrogen and helium gas, both gas and dust 
travelling with a speed of $\rm 26\,km\,s^{-1}$
\cite{gruen1994a,baguhl1995a,witte1996,frisch1999}. 
The interstellar dust flow persisted at high latitudes above and below
the ecliptic plane and even over the poles of the Sun, whereas 
interplanetary dust was strongly depleted at high ecliptic latitudes.

Later measurements with the Galileo dust detector in the ecliptic plane 
confirmed the Ulysses results:
beyond about 3~AU the interstellar dust flux exceeds the flux of
micron-sized interplanetary grains. Furthermore, interstellar dust
is ubiquitous in the solar system: dust measurements between 0.3 
and 3 AU in the ecliptic plane exist also from Helios, Galileo 
and Cassini. This data shows evidence for distance-dependent 
alteration of the interstellar dust stream caused by radiation 
pressure, gravitional focussing and electromagnetic interaction 
with the time-varying interplanetary magnetic field which also
depends on grain size 
\cite{altobelli2003,altobelli2005a,altobelli2005b,mann2000a,landgraf2000b,czechowski2003}.
As
a result, the size distribution and fluxes of grains measured inside the 
heliosphere are strongly modified \cite{landgraf1999a,landgraf2003}.

Interstellar grains observed with the spacecraft detectors range 
from $\rm 10^{-18}\,kg$ to above $\rm 10^{-13}\,kg$. If we compare the
mass distribution of these interstellar impactors detected in-situ with
the dust mass distribution derived from astronomical observations, we
find that the in-situ measurements overlap only with the largest 
masses observed by remote sensing. It indicates that
the intrinsic size distribution of interstellar grains in the LIC
extends to grain sizes larger than those detectable by astronomical 
observations \cite{frisch1999,frisch2003,landgraf2000a,gruen2000b}. 
Even bigger interstellar
grains (above $\rm 10^{-10}\,kg$) are observed as
radar meteors entering the Earth's atmosphere \cite{taylor1996b,baggaley2002}. 
The flow direction of these larger grains varies over a much wider angular
range than that of small particles measured by the in-situ detectors.

The total grain mass detected in-situ by Ulysses, which includes bigger grains
in the LIC than those detectable with astronomical techniques, led to
the conclusion that the LIC gas is enhanced with the refractory
elements (e.g. Fe, Mg, Mn) that would ordinarily dominate the mass of
interstellar dust grains. Earlier investigations indicated an enhancement of 
the dust-to-gas mass ratio in the LIC by up to a factor of five 
\cite{frisch1999}. Recent reanalysis with improved solar heavy element 
abundances and an updated value for the sensitive area of the Ulysses 
dust detector \cite{altobelli2004} brought this enhancement down to 
a factor of two \cite{frisch2003}(Slavin \& Frisch, this volume).

\begin{figure}
\hspace{-0.2cm}
\centerline{\includegraphics[clip=,width=1.04\textwidth]{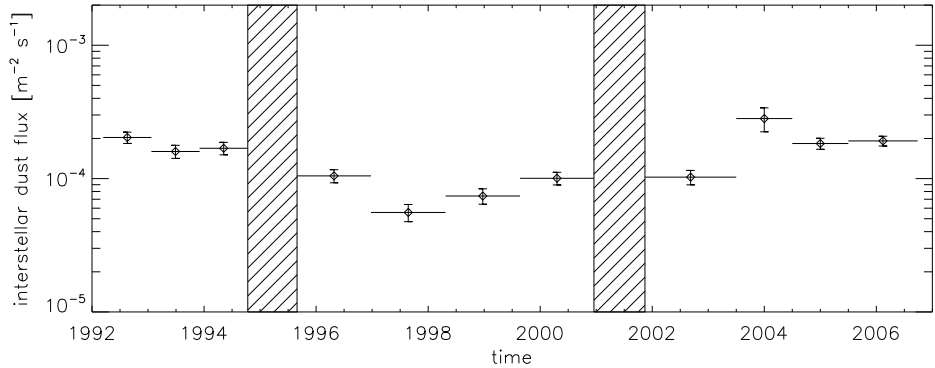}}
         \caption{\label{flux}
Interstellar dust flux measured by Ulysses. The horizontal lines indicate 
the length of the time intervals, and the vertical bars of the data points 
represent the $\rm 1\,\sigma$ uncertainty due to small number statistics. The
dashed regions in 1995 and 2001 show the periods of Ulysses' perihelion passages
where the distinction of interstellar dust from interplanetary impactors is
difficult. Furthermore, in 2003 and 2004 a possible contamination by jovian dust 
stream particles around Jupiter flyby, which occurred in February 2004, may lead 
to an erroneously enhanced 
interstellar flux. Dust streams identified in the Ulysses data set 
were removed by ignoring the time interval when a dust stream occurred.
}
\end{figure}

In addition to the distribution of grain masses, the instrument has 
monitored the flux of interstellar dust particles through the
heliosphere since Ulysses left the ecliptic plane in 1992 (Figure~\ref{flux}). 
In mid 1996, a drop of the interstellar dust flux from initially $1.5\times 10^{-4}\:{\rm
m}^{-2}\:{\rm s}^{-1}$ to $0.5\times 10^{-4}\:{\rm m}^{-2}\:{\rm
s}^{-1}$ occurred  \cite{landgraf1999b}. 
Since early 2000, Ulysses has detected interstellar dust flux levels 
above $10^{-4}\:{\rm m}^{-2}\:{\rm s}^{-1}$ again. The drop in 1996
was explained by
increased filtering of small grains by the solar wind driven magnetic field
during solar minimum conditions \cite{landgraf2000a,landgraf2000b}. 
The filtration 
caused a deficiency of detected interstellar grains with sizes below 
$0.2\:{\rm \mu m}$~\cite{gruen1994a}. An additional filtration by solar 
radiation pressure, which was found to be effective at heliocentric distances 
below $4\:{\rm AU}$, deflects grains with sizes of $0.4\:{\rm \mu m}$ 
\cite{landgraf1999a}.

Modelling the dynamics of the electrically 
charged dust grains in the heliosphere can give us information 
about the Local Interstellar Cloud (LIC) where the particles 
originate from. In the time interval 2001 to 2003 the interstellar dust 
flux stayed relatively constant, in agreement with improved models 
\cite{landgraf2003}. 
The dominant contribution to the flux comes from grains with a 
charge to mass ratio $q/m = \rm 0.59\,C\,kg^{-1}$ and a radiation
pressure efficiency of $\beta = 1.1$
which -- in the simulation -- corresponds to a grain radius 
of $0.3\,\mu\mathrm{m}$ (assuming spherical 
grains). The models assume a constant dust concentration in the 
LIC and give a good fit to the dust fluxes  measured 
between end-1992 and end-2003. The fact that the models fit the
observed variations implies that the dust phase of the LIC
is homogeneously distributed over length scales of at least 50~AU which
is the distance inside the LIC traversed by the Sun during this time 
period. This result, however, needs to be reexamined in light of
Ulysses's most recent measurements obtained during the spacecraft's 
$\mathrm{3^{rd}}$ passage through the outer solar system. 

\section{Ulysses's $\mathbf{3^{rd}}$ passage through the outer solar system}

\label{sec_meas}

From 2002 to 2006 Ulysses made its $\mathrm{3^{rd}}$ passage through the outer
heliosphere (aphelion passage occurred in June 2004 at a heliocentric 
distance of 5.4~AU), providing again good conditions for measuring interstellar dust.
In February 2004, however, the 
spacecraft had its second Jupiter flyby at a closest approach distance
0.8~AU which also allowed for good measurement conditions for the
dust streams emanating from the jovian system
\cite{gruen1993,krueger2006c}. A total of 28 streams were detected,
more than twice the number of detections from Ulysses' first Jupiter
flyby in 1992: the first stream was recorded 
in November 2002 when Ulysses was still 3.4~AU 
away from Jupiter, and the last stream in 
mid-2005 at more than 4~AU jovicentric distance.

In the outer solar system and at high ecliptic latitudes the
interstellar impactors can usually be identified by their
impact direction: they approach from a retrograde
direction while the majority of interplanetary grains move on 
prograde heliocentric orbits. During most of the time in the
2002 to 2005 interval, however, the jovian dust streams 
approached from roughly the same direction \cite{krueger2007}, so that
identification of interstellar grains by their impact direction
alone was not possible. On the other hand, the measured impact 
charge distribution showed that the majority of grains with 
impact charges above $Q_{\rm I} = 2 \times 10^{-13}\,\rm C$ are 
of interstellar origin while most jovian stream
particles have impact charges below this limit.
We therefore use this limit to separate both populations 
of impactors.
Contamination by jovian stream particles, however,
cannot entirely be excluded this way, and in particular the fluxes
attributed to interstellar grains in 2004 --
when the instrument detected the most intensive 
jovian dust streams \cite{krueger2006c} -- may be contaminated 
by jovian impactors. Hence, the data point
in Figure~\ref{flux} showing an elevated interstellar flux in 2004
has to be taken with caution. 
On the other hand, in mid-2005 dust stream detections
ceased, and in the later data the contribution by
jovian stream particles should be negligible.

\begin{figure}
\vspace{-3.8cm}
\hspace{-0.3cm}
\begin{turn}{90}
\centerline{\includegraphics[clip=,width=0.37\textwidth]{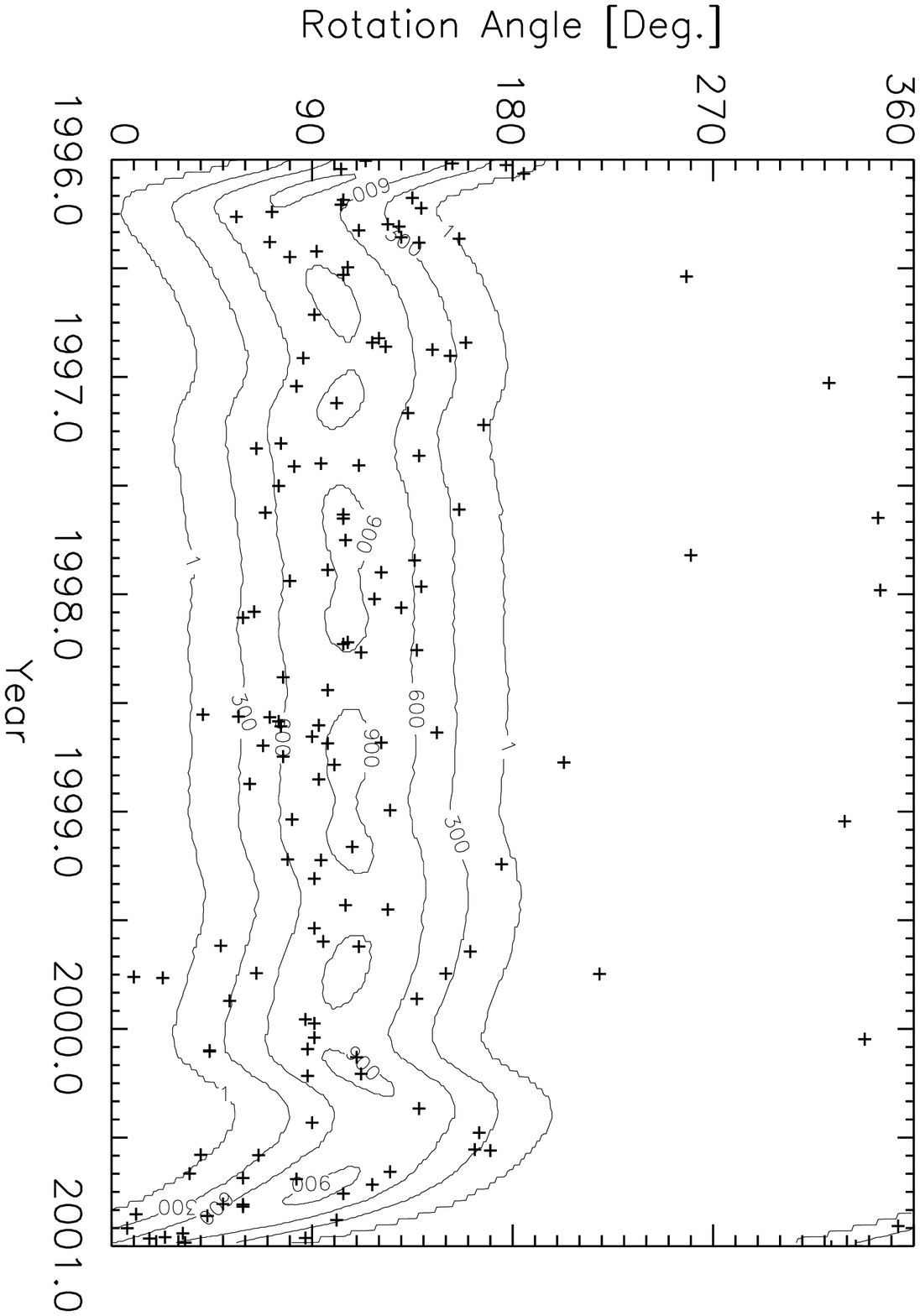}}
\end{turn}
\vspace{-3.8cm}
\begin{turn}{90}
\centerline{\includegraphics[clip=,width=0.37\textwidth]{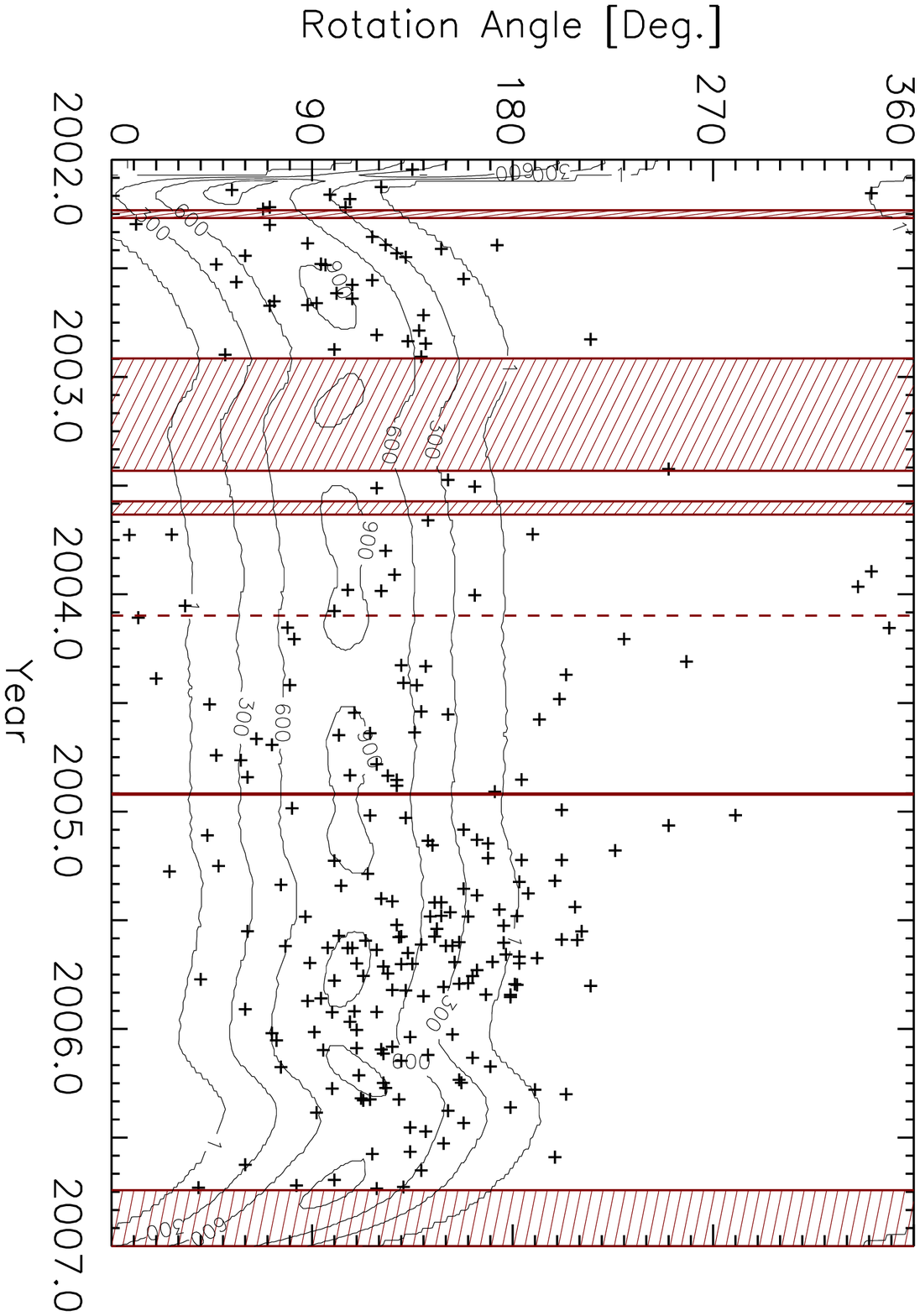}}
\end{turn}
         \caption{\label{rot}
Impact direction (i.e. spacecraft rotation angle at dust particle impact) 
of interstellar grains measured with Ulysses in two time intervals. 
{\em Left:} 1 January 1996 to 31 December 2000;
{\em right:} 1 January 2002 to 31 December 2006.
Ecliptic north is close to $0^{\circ}$; impact charges 
$\QI \geq 2\times 10^{-13}\,\mathrm{C}$.
Each cross indicates an individual impact.
Contour lines show the effective sensor area for particles
approaching from the upstream direction of interstellar 
helium. In the right panel, a vertical dashed line shows Jupiter 
closest approach on 5 February 2004, five shaded areas indicate 
periods when the dust instrument was switched off.
}
\end{figure}

Figure~\ref{rot} shows the impact directions of grains 
with impact charges  $\QI \geq 2 \times 10^{-13}\,\mathrm{C}$ 
for two different time intervals. The intervals were chosen 
such that Ulysses was traversing approximately
the same region of the outer solar system during both periods.
Contour lines show the effective dust sensor area for particles
approaching from the upstream direction of interstellar helium
\cite{witte1996},
implying that the detection conditions for 
interstellar dust were very similar in both intervals.
The approach directions of the majority of grains are consistent 
with the upstream direction of the interstellar helium flow.
This is particularly evident in the
earlier time interval 1996 to 2000 
({\em left panel} in Figure ~\ref{rot}). It should be noted 
that in this 
interval Jupiter and Ulysses were on opposite sides
of the solar system, separated by more than 10~AU,
so that contributions by jovian stream particles can be excluded. 
The distribution of the measured 
rotation angles is also shown in Figure~4. In the 1997 to 1999 interval
the average impact direction of the interstellar grains
was at rotation angles of about $95^{\circ}$. 

The interstellar impactors were still concentrated towards
the interstellar helium flow direction in 2003 and 2004 
({\em right panel} in Figure~\ref{rot}) although the 
distribution of the measured rotation angles was
somewhat wider. Later, in 2005, the impact directions 
were significantly shifted from the helium flow. 
This is also evident in the {\em right panel} of Figure~4:
the mean rotation angle of the impactors
is at $135^{\circ}$ rotation angle. Taking into account that the
detection geometry has slighly changed between the two time 
intervals, this implies that the interstellar dust flow has
shifted by at least $30^{\circ}$ in southward direction, away 
from the ecliptic plane. The wider distribution of impact directions
is also evident. 

\begin{figure}
\vspace{-3.8cm}
\hspace{-0.5cm}
\begin{turn}{90}
\centerline{\includegraphics[clip=,width=0.39\textwidth]{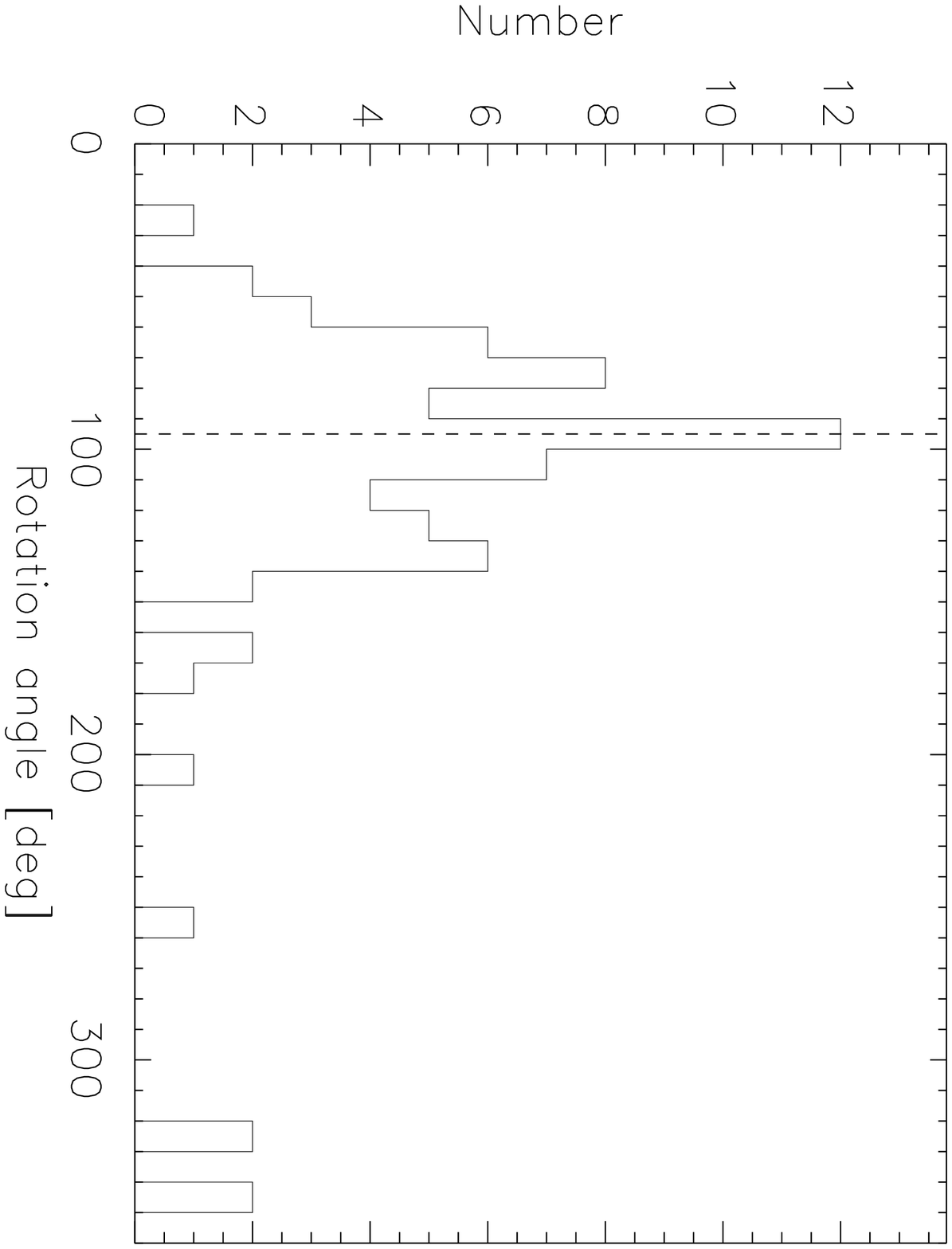}}
\end{turn}
\vspace{-3.8cm}
\hspace{-0.4cm}
\begin{turn}{90}
\centerline{\includegraphics[clip=,width=0.39\textwidth]{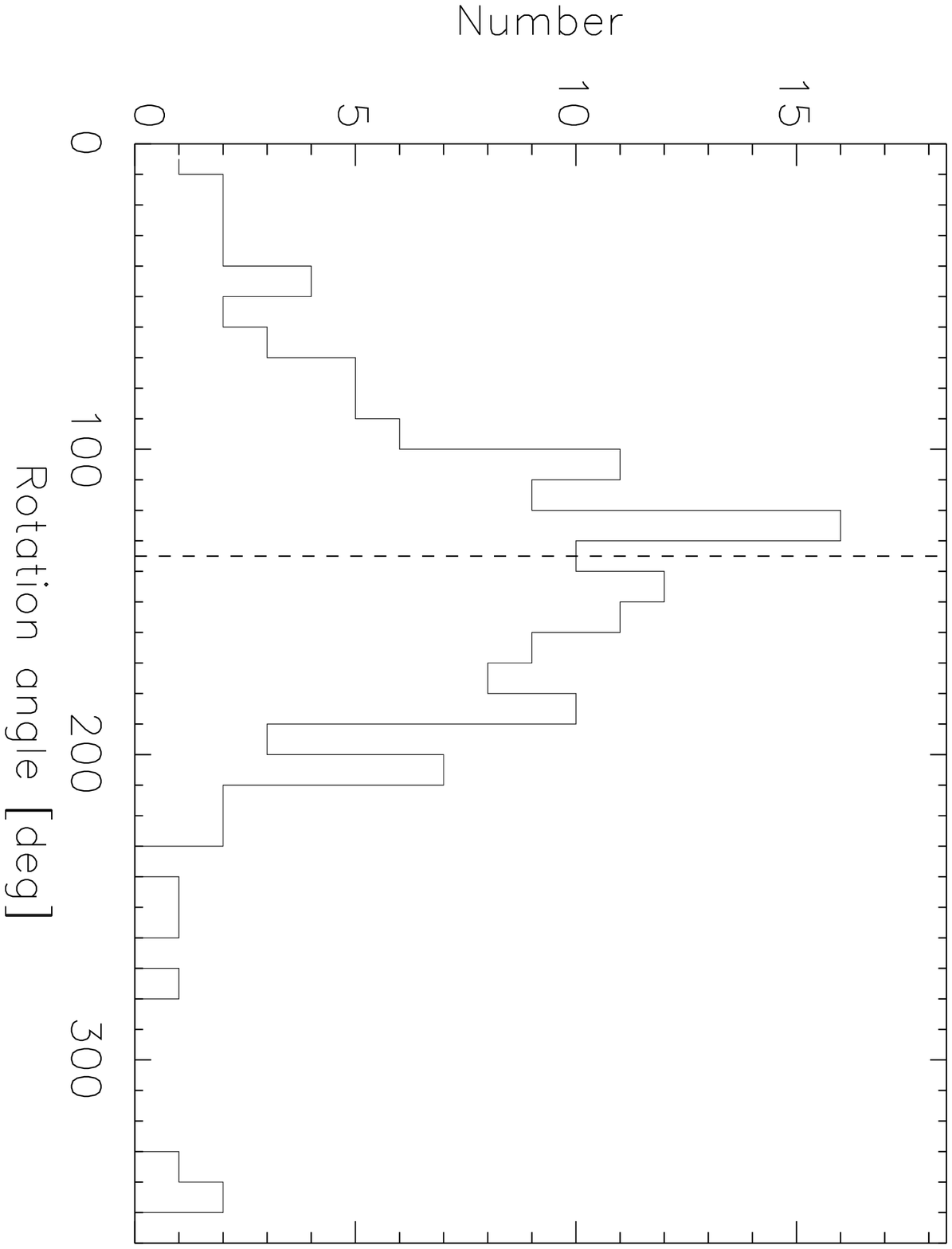}}
\end{turn}
         \caption{\label{distribution}
Distribution of measured impact directions (i.e. spacecraft rotation angle at 
dust particle impact) of interstellar impactors for two time intervals. 
{\em Left:} 1 January 1997 to 31 December 1999; {\em right:} 1 January 2003 to 31 December 2005.
In the earlier time interval the maximum of the distribution is at a rotation angle 
of $95^{\circ}$ very close to the value expected from the interstellar helium flow.
In the second interval the maximum is at $135^{\circ}$. 
}
\end{figure}

\section{Discussion}

\label{sec_discussion}

The dust measurements from Ulysses' $\mathrm{3^{rd}}$ passage through the outer
solar system imply an at least $30^{\circ}$
shift in the approach direction of interstellar dust grains.
The reason for this shift remains mysterious. Whether it is connected 
to a secondary stream of interstellar neutral atoms shifted from 
the main neutral gas flow \cite{collier2004} is presently unclear. 
However, given that the neutral gas stream is shifted along the ecliptic 
plane while the shift in the dust flow is offset from the ecliptic, 
a connection between both phenomena seems unlikely.

Even though Ulysses' position in the heliosphere and the dust
detection conditions were very similar during both time intervals
considered here, the configurations of the solar wind driven 
interplanetary magnetic
field (IMF) which affected the grain dynamics were vastly different. 
One has to take into account that the interstellar grains need 
approximately  
twenty years to travel from the heliospheric boundary to the inner 
solar system where they are detected by Ulysses. Thus, the effect 
of the IMF on the grain dynamics is the accumulated effect caused
by the interaction with the IMF over several years. 
Hence, in the earlier time interval (1997-1999) the grains had a 
recent dynamic history dominated by solar minimum conditions
\cite{landgraf2000b},
while the grains detected during the second interval (2002-2005) 
had a recent history dominated by the much more disturbed solar maximum 
conditions of the IMF. This latter configuration 
may have a strong influence on the dust dynamics in the inner heliosphere
but it is not modeled in detail in the presently existing models. 
It may particularly affect 
small grains which are most sensitive to the electromagnetic 
interaction. One would expect a size-dependent shift in the grain 
impact direction which, however, is not evident in the data. Whether these 
phenomena cause the observed shift in the 
approach direction of the interstellar dust grains will be the 
subject of future investigations. 


\acknowledgements 
We thank the Ulysses project at ESA and NASA/JPL for 
effective and successful mission operations. 
This work has been supported by the Deutsches Zentrum f\"ur 
Luft- und Raumfahrt e.V. (DLR) under grants 50 0N 9107 and 50 QJ 9503. 
Support by Max-Planck-Institut f\"ur Kernphysik and Max-Planck-Institut f\"ur 
Sonnensystemforschung is also gratefully acknowledged.

\bibliographystyle{aa}


\end{article}
\end{document}